\documentclass[a4paper,11pt,nofootinbib]{revtex4}
\pdfoutput=1

\usepackage{graphicx}
\usepackage{hyperref}
\usepackage{amssymb}
\usepackage{latexsym}
\usepackage{multirow}
\usepackage{amsmath}
\usepackage{mathrsfs}  
\usepackage{xcolor}

\newcommand{\AddrUNAM}{{\it Instituto de F\'{\i}sica, Universidad Nacional Aut\'onoma de M\'exico, A.P. 20-364, Ciudad de M\'exico 01000, M\'exico.}}
\newcommand{\AddrNapoliA}{
  {\it Dipartimento di Fisica {\it Ettore Pancini}, Universit\`a di Napoli Federico~II, Complesso Univ. Monte S. Angelo, I-80126 Napoli, Italy}}
    \newcommand{\AddrNapoliB}{{\it INFN, Sezione di Napoli, Complesso Univ. Monte S. Angelo, I-80126 Napoli, Italy}}
  \newcommand{\AddrSouthampton}{
  {\it Physics and Astronomy, University of Southampton, Southampton, SO17 1BJ, U.K.}}

\makeatletter

\renewcommand*{\p@subsection}{}

\renewcommand*{\p@subsubsection}{}
\makeatother

\begin{document}

\title{Neutrinophilic Dark Matter in the epoch of IceCube and Fermi-LAT}

\author{Marco~Chianese}
\email{ma.chianese@gmail.com}\affiliation{\AddrSouthampton}
\author{Gennaro~Miele}
\email{miele@na.infn.it}\affiliation{\AddrNapoliA}\affiliation{\AddrNapoliB}
\author{Stefano~Morisi}
\email{stefano.morisi@gmail.com} \affiliation{\AddrNapoliA}\affiliation{\AddrNapoliB} 
\author{Eduardo~Peinado}
\email{epeinado@fisica.unam.mx}\affiliation{\AddrUNAM}

\begin{abstract}
The recent observation of the blazar TXS~0506+056 suggests the presence of a hard power-law component in the extraterrestrial TeV-PeV neutrino flux, in agreement with the IceCube analysis on the 8-year through-going muon neutrinos from the Northern Sky. This is slightly in tension with the soft power-law neutrino flux deduced by the IceCube 6-year High Energy Starting Events data. A possible solution to such a puzzle is assuming a two-component neutrino flux. In this paper, we focus on the case where, in addition to an astrophysical power-law, the second component is a pure neutrino line produced by decaying Dark Matter particles. We investigate how to realize a neutrinophilic decaying Dark Matter in an extension of the Standard Model. The main features of the model are: i) the requirement of a new symmetry like a global $U(1)$ charge; ii) the Dirac nature of active neutrinos; iii) a low-reheating temperature of the Universe of about 1~TeV. We perform a likelihood statistical analysis to fit the IceCube data according to the present Fermi-LAT gamma-rays constraints.
\end{abstract}

\maketitle

\newpage
\tableofcontents

\section{Introduction}

The IceCube Collaboration detected a flux of high energy neutrinos (with energy above 60 TeV) whose origin must be astrophysical, namely the observed flux cannot be explained in terms of atmospheric neutrinos. The fit of the 6-year High Energy Starting Events (HESE) data provides a power-law flux with spectral index $2.92^{+0.29}_{-0.33}$~\cite{Aartsen:2017mau}. On the other hand, an analysis of the 8-year through-going muon neutrinos from Northern Sky, with energy bigger than 200~TeV, yields a spectral index  $2.19\pm 0.10$~\cite{Aartsen:2017mau}. Remarkably, the recent measurement of the coincident neutrino and gamma-ray emission from the blazar TXS~0506+056~\cite{IceCube:2018cha,Ahnen:2018mvi} also confirms a best-fit spectral index in the range $2.0\div2.3$ depending on the data sets considered and fit procedures assumed~\cite{IceCube:2018dnn}. A spectral index $\gamma \simeq 2.2$ is slightly in tension with the 6-year HESE data (about 2$\sigma$) and with the global analysis of all available IceCube data (about 3$\sigma$)~\cite{Aartsen:2015knd}. Such a discrepancy among different IceCube data samples could be just a statistical fluctuation or could indicate the presence of different components in the extraterrestrial neutrino flux. This consideration has pushed the scientific community to investigate a two-component neutrino flux scenario~\cite{Chen:2014gxa,Palladino:2016zoe,Vincent:2016nut,Palladino:2016xsy,Anchordoqui:2016ewn,Palladino:2017qda,Palladino:2018evm,Sui:2018bbh}. The IceCube observations are indeed consistent with a hard isotropic extragalactic neutrino flux together with an additional softer component having a potential Galatic origin~\cite{Aartsen:2017mau,Aartsen:2015knd} (giving a dominant contribution to the Southern Sky). Remarkably, as pointed out in Ref.~\cite{Chianese:2017jfa}, the tension of the diffuse neutrino flux with the assumption of a single power-law is strengthened once the 6-year HESE data are combined with the latest 9-year ANTARES data~\cite{Albert:2017bdv}. Moreover, both IceCube and ANTARES experiments show a slightly excess in the same energy range $\left(40-200~{\rm TeV}\right)$ once a power-law flux with spectral index $\gamma \leq 2.2$ is considered~\cite{Chianese:2016opp,Chianese:2016kpu,Chianese:2017jfa,Chianese:2017nwe}.

Hence, so far the origin of the observed TeV-PeV astrophysical neutrino flux is unclear. At the same time the searches for spatial and temporal correlations with gamma-rays pose strong constraints to several extragalactic astrophysical candidates, providing that they can have only a sub-dominant contribution to extraterrestrial neutrino flux~\cite{Bechtol:2015uqb,Aartsen:2016lir,Aartsen:2017wea}.\footnote{We note that it has recently pointed out that the blazar TXS~0506+056 could account for 1\% of the diffuse TeV-PeV neutrino flux~\cite{IceCube:2018dnn}.} In addition to the standard astrophysical sources, it has been proposed the existence of hidden astrophysical sources that do not have a gamma-ray counterpart~\cite{Kimura:2014jba,Murase:2015xka,Senno:2015tsn}.

Heavy Dark Matter particles could also produce high-energy neutrinos through their decay~\cite{Anisimov:2008gg,Feldstein:2013kka,Esmaili:2013gha,Bai:2013nga,Ema:2013nda,Esmaili:2014rma,Bhattacharya:2014vwa, Higaki:2014dwa,Rott:2014kfa,Ema:2014ufa, Murase:2015gea,Dudas:2014bca,Fong:2014bsa,Aisati:2015vma,Ko:2015nma,Dev:2016qbd,Fiorentin:2016avj,DiBari:2016guw,Zavala:2014dla,Anchordoqui:2015lqa,Dev:2016uxj,Chianese:2016smc,Borah:2017xgm,Boucenna:2015tra,Chianese:2016opp,Chianese:2016kpu,Hiroshima:2017hmy,Bhattacharya:2017jaw,ElAisati:2017ppn,Bhattacharya:2014yha,Bhattacharya:2016tma,Chianese:2017nwe,Kachelriess:2018rty,Sui:2018bbh,Dhuria:2017ihq}. A decaying Dark Matter candidate with mass of about 100 TeV could alleviate the tension between the HESE and through-going muon neutrino data samples~\cite{Chianese:2016opp,Chianese:2016kpu,Chianese:2017nwe}. On the other hand, annihilating Dark Matter is not a viable scenario since the interpretation of the neutrino flux would require too large cross-sections that are in general not allowed by unitarity~\cite{Chianese:2016kpu,Feldstein:2013kka}. Depending on the specific Dark Matter decay channel, namely on the particular model, there is also the production of charged particles and gamma-rays. In Ref.~\cite{Cohen:2016uyg} it has been shown that most of such Dark Matter models, especially the ones with hadronic final states (see also Ref.s~\cite{Chianese:2016kpu,Chianese:2017nwe}), are excluded or in tension with limits coming from gamma-ray Fermi-LAT data~\cite{Ackermann:2014usa}. In particular, the most favorable case is a Dark Matter decaying only into neutrinos.

In this paper, we study in more detail the production of a neutrino line from Dark Matter decay from the model building point of view and investigate how such a neutrinophilic Dark Matter can be produced in the early Universe. Then, we perform a fit of the extraterrestrial TeV-PeV neutrino flux observed by IceCube after 6-year of data taking.\footnote{Here, we prefer to consider the 6-year HESE data sample since the new 7.5-year HESE data sample recently presented by IceCube during the conference Neutrino 2018 is preliminary and under further investigation. Moreover, we expect that such new data are not going to change substantially our conclusions.} Motivated by the recent measurements related to the blazar TXS~0506+056 and by the IceCube analysis of through-going muon neutrinos, we consider an astrophysical neutrino flux with spectral index 2.2 as benchmark. Hence, we provide the allowed regions of the parameter space of the model in agreement with neutrinos and gamma-rays observations.

The paper is organized as follows. In Section~2 we describe the model that provides a neutrinophilic decaying Dark Matter. In Section~3 we discuss how such a heavy Dark Matter candidate can be produced in the early Universe. In Section~4 we analyze the compatibility of the model with the recent IceCube and Fermi-LAT data. Finally, in Section~5 we draw our conclusions.

\section{The model}

According to the Table S2 of Ref.~\cite{Cohen:2016uyg}, the only renormalizable operator for a Dark Matter neutrino line is obtained by extending the Standard Model with a scalar $SU(2)_L$-triplet with hypercharge $Y=+1$
\begin{equation}
\Delta=\sum_{i=1}^{3}\delta_i\tau_i=\left(\begin{array}{cc}\Delta^+&\sqrt{2}\Delta^{++}\\\sqrt{2}\Delta^0
&-\Delta^+\end{array}\right)\,,
\end{equation}
where $\Delta^0 \equiv \frac{1}{\sqrt{2}} \left( \delta_1 + i\delta_2 \right)$, $\Delta^+ \equiv \delta_3$, $\Delta^{++} \equiv \frac{1}{\sqrt{2}} \left( \delta_1 - i\delta_2 \right)$ and $\tau_i$ are the Pauli matrices. In this way the Standard Model Lagrangian is extended with new physics terms given by
\begin{equation}
\mathcal{L} \supset \mathcal{L}_{kin} + \mathcal{L}_{\nu}+V\,,
\end{equation}
where $\mathcal{L}_{kin}$ is the kinetic term for the scalar triplet, $V$ is the scalar potential involving the Higgs field $H$, and
\begin{equation}\label{Ltripletnu}
\mathcal{L}_{\nu} = 
\frac{1}{2} \lambda_{ij} L_i^T C^{-1} i\tau_2 \Delta \,L_j+\mbox{h.c.} \,,
\end{equation}
where 
\begin{equation}
L_i=\left(\begin{array}{c}\nu_{i L}\\\ell^-_{i L}\end{array}\right)
\end{equation}
is the lepton left-handed doublet and $\lambda_{ij}$ is a complex symmetric matrix. The components of $\Delta$ and $H$ are complex fields, they can be defined by an expansion around the corresponding v.e.v.'s as
\begin{equation}\label{expansion}
\begin{array}{l l l}
H^0 = v+ h^0 +i\, G^0 \,, \qquad & H^+ = G_1 +i\, G_2 \,, & \\
\Delta^0 =v_\Delta + \eta^0 + i \, A^0  \,,\qquad & \Delta^+ =A_1 +i A_2\,, \qquad &  \Delta^{++} = A_3+i A_4\,,
\end{array}
\end{equation}
where $v=174$~GeV is the Higgs v.e.v.. Then, by substituting \eqref{expansion} in the expression \eqref{Ltripletnu} we get the coupling
\begin{equation}\label{DMdecaynu}
\frac{1}{\sqrt2}\lambda_{ij} \,\chi\, \nu^T_{iL}C^{-1}\nu_{jL}+\mbox{h.c.}\,,\end{equation}
that is responsible for the decay of the Dark Matter into a neutrino couple, $\chi\to\overline{\nu_i}\,\overline{\nu_j}$ $\left(\chi^*\to\nu_i\,\nu_j\right)$, where we have defined the Dark Matter field as $\chi\equiv \eta^0+i A^0$. The structure of the matrix $\lambda$ defines the branching ratios of Dark Matter decay channels into different neutrino flavors. In this paper, for the sake of simplicity we study the case $\lambda_{ij} = \delta_{ij} \lambda$\footnote{This is the case for instance in a framework of the $A_4$ flavor symmetry when the left-handed leptons $L$ transform as a triplet under the flavor symmetry~\cite{Bonilla:2017ekt}.} that provides for the total Dark Matter neutrino flux the flavor ratio $\left(f_e:f_\mu:f_\tau\right)=\left(\frac13:\frac13:\frac13\right)$.

We note that the scalar $\Delta$ couples to fermionic matter only {\it via} the terms of Eq.~\eqref{Ltripletnu}, and it does not couple with quarks because of the $SU(3)_C$ color conservation.  The triplet $\Delta$  interacts also with gauge bosons through the kinetic term, and with the Higgs doublet through the scalar potential
\begin{eqnarray}\label{Vtriplet}
V &=& -\mu^2H^\dagger H +\lambda_0(H^\dagger H)^2 + M^2 \mbox{Tr}( \Delta^\dagger \Delta) +\lambda_1H^\dagger H  \mbox{Tr}( \Delta^\dagger \Delta )+ \lambda_2  \left[\mbox{Tr}\, (\Delta^\dagger \Delta)\right]^2 + \nonumber\\
&+&\lambda_3 \mbox{Tr}\, (\Delta^\dagger \Delta \Delta^\dagger \Delta)+
\lambda_4H^\dagger \Delta \Delta^\dagger H+\mu_\Delta H^\dagger \Delta \tilde{H} +\mbox{h.c.}\,,
\end{eqnarray}
where $\tilde{H} = i\tau_2 H^*$, $\lambda_i$ are dimensionless couplings and $\mu$, $\mu_\Delta$ are couplings with dimension of a mass. The terms proportional to $\lambda_1$, $\lambda_4$ and $\mu_\Delta$ mix $\Delta$ with $H$ and can lead to additional Dark Matter decay channels. In particular, the term proportional to $\mu_\Delta$ induces the decay of the Dark Matter particles into two Higgs scalars. Moreover, since the v.e.v. of $\Delta$ is proportional to $\mu_\Delta$, other Dark Matter decay channels are induced by the terms with couplings $\lambda_1$ and $\lambda_4$ through $\mu_\Delta$.

Therefore, in order to have only decays into neutrinos, we require $\mu_\Delta=0$ in the scalar potential $V$ or we have to impose a symmetry that forbids such a coupling. An example of such a symmetry is a global $U(1)_L$ symmetry under which only leptons and $\Delta$ are charged.\footnote{It is possible under certain assumptions that when a continuous global symmetries is imposed Planck-suppressed non-renormalizable operators could also induce Dark Matter decay channels but this applies when the Dark Matter is a singlet under $SU(2)_L$~\cite{Mambrini:2015sia}. Our model is also compatible when considering instead of $U(1)_L$ a $Z_n$ symmetry with $n\geq3$.} The charge assignments are reported in Table~\ref{tab1} where $\ell_R$ denotes the right-handed charged leptons and $\nu_R$ denotes the right-handed neutrinos. We note that, if $\mathcal{Q}_L(\Delta)=0$, then $U(1)_L$ would correspond to the Standard Model lepton number. However, differently from the Standard Model where the lepton number is an accidental symmetry,  here it is assumed as a global symmetry of the Lagrangian. In summary, we impose by hand the lepton number, we extend the Standard Model with a scalar electroweak triplet (and with right-handed neutrinos) and we charge such a scalar with respect to the lepton number itself. Since the neutral component $\chi$ of $\Delta$ is the Dark Matter candidate, also the Dark Matter is charged under such an extended lepton number.
\begin{table}[t!]
\begin{center}
\begin{tabular}{|c|c|c|c|c|c|}
\hline
& $L$ & $\ell_R$& $H$&  $\Delta$ &$\nu_R$\\
\hline
$SU(2)_L$ & 2 & 1&2 & 3 & $1$\\\hline
$U(1)_Y$ & -1/2 & -1&1/2 & 1 & 0\\
\hline
$U(1)_L$ & $1$ & $1$ & $0$  & $-2$ & $1$  \\
\hline
\end{tabular}\caption{Relevant particle content and quantum numbers of the model. Quarks are not reported being not charged under $U(1)_L$ and not used here.} \label{tab1}
\end{center}
\end{table}

It is worth observing that since the field $\Delta$ does not acquire a v.e.v., the operator in Eq.~\eqref{Ltripletnu} does not generate a mass to the active light neutrinos as in the standard type-II seesaw mechanism~\cite{Schechter:1980gr,Lazarides:1980nt,Mohapatra:1980yp}. Since the $U(1)_L$ charge must remain preserved to have the Dark Matter neutrino line only, as it is well known, there is no way to generate a Majorana mass term for neutrinos because it would violate the $U(1)_L$ symmetry. Therefore, the only way to generate a neutrino mass is by introducing right-handed neutrino fields $\nu_R$ with quantum numbers as reported in Tab.~\ref{tab1} and allowing for the Dirac Yukawa interaction
\begin{equation}
y_{ij} \overline{L_i} \tilde{H} \nu_{Rj} +\mbox{h.c.} \,.
\end{equation}
Such a neutrino Yukawa coupling is related to neutrino masses and mixing parameters~\cite{deSalas:2017kay,Capozzi:2018ubv,Esteban:2016qun}. In general, the two couplings $\lambda_{ij}$ and $y_{ij}$ are independent on each other, but if one assumes a flavor symmetry to induce a pattern to the neutrino Yukawa matrix, this would impose a particular structure to the matrix $\lambda$ as well~\cite{Aranda:2013gga,Bonilla:2017ekt} (see Ref.~\cite{King:2014nza} for a review about flavor symmetries). Hence, a flavor symmetry in the neutrino sector would provide specific branching ratios in the Dark Matter decays and, consequently, a particular flavor ratio $\left(f_e:f_\mu:f_\tau\right)$ to the Dark Matter neutrino flux. We note that different flavor ratios at production could be experimentally discriminated by neutrino telescopes in the near future~\cite{Mena:2014sja,Palladino:2015zua,Arguelles:2015dca,Bustamante:2015waa}.

Let us now study the masses and the couplings of the physical fields by inserting in the scalar potential and in the kinetic term the expansions given in Eq.~\eqref{expansion}. Since $\Delta$ does not pick up a v.e.v. , the study of the scalar potential is simplified providing that the neutral component $G^0$ does not mix with $A^0$. For the same reason, the components $A_{1,2,3,4}$ of the triplet do not mix with with $G_{1,2}$ of the Higgs doublet. Therefore, the three fields $G^0,\,G^\pm$ are the usual Goldstone bosons of the Standard Model. On the other hand, all the components of the scalar triplet $\Delta$ are physical and their masses are given by\footnote{Note that the scalar and pseudoscalar components are degenerate in mass and both are Dark Matter candidates.}
\begin{eqnarray}\label{masstriplet}
m_{\chi}^2&=&M^2+\frac12\left(v^2 \, \lambda_1+ v^2 \, \lambda_4\right) \,,\\
m_{\Delta^+}^2&=&M^2+\frac{1}{2}v^2 \, \lambda_1+\frac{1}{4}v^2 \, \lambda_4 \,,\\
m_{\Delta^{++}}^2&=&M^2+\frac{1}{2}v^2 \, \lambda_1\,,
\end{eqnarray}
in terms of the couplings of the scalar potential. In other words, given the Dark Matter mass $m_\chi$ and the squared mass splitting $\Delta m^2 = - v^2 \lambda_4/4$, the masses of the charged components can be cast as
\begin{eqnarray}
m_{\Delta^+}^2 = m_\chi^2 + \Delta m^2 & \qquad{\rm and}\qquad & m_{\Delta^{++}}^2 = m_\chi^2 + 2 \, \Delta m^2\,.
\end{eqnarray}
Therefore, for $\Delta m^2 >0$ the neutral component $\chi$ is the lightest one and, consequently, its decays into the others components of the triplet are not kinematically allowed.

Finally, the interactions of the Dark Matter particles with scalars and gauge bosons can be summarized as follows:
\begin{itemize}
\item three scalars: \\ $\chi H^+\Delta^-$,~~$|\chi|^2 h_0$
\item four scalars: \\ $\chi H^+\Delta^- h_0$,~~$\chi \Delta^{++}\Delta^-\Delta^-$,~~$|\chi|^2 H^+H^-$,~~$|\chi|^2 h_0^2$,~~$|\chi|^2 \Delta^+\Delta^-$,~~$|\chi|^2 \Delta^{++}\Delta^{--}$
\item two scalars and one vector boson: \\ $\chi W^+_\mu\gamma^\mu\Delta^-$,~~$\left(\partial_\mu \chi^*\, \chi-\chi^*\partial_\mu \chi\right) Z^\mu$
\item 2 scalars and 2 vector bosons:  \\ $\chi W^+_\mu\gamma^\mu \Delta^-$,~~$\chi W^+W^+\Delta^{--}$,~~$\chi W^+ Z \Delta^-$,~~$|\chi|^2 W^+W^-$,~~$|\chi|^2ZZ$
\end{itemize}
and the corresponding hermitian conjugates. From these couplings, one can directly read all annihilation and co-annihilation channels.

\section{Dark Matter production in the early Universe}

As discussed in the previous Section, our Dark Matter candidate has weak interactions and it is therefore a Weakly Interacting Massive Particle (WIMP). Indeed, in the early Universe the main processes responsible for the Dark Matter production are the annihilations involving $SU(2)_L\otimes U(1)_Y$ vector bosons, whose $s$-wave term of the thermally averaged total cross-section takes the expression~\cite{Cirelli:2005uq}
\begin{equation}
\left< \sigma v \right> = \frac{3\,g_2^4 + 4\,g_2^2 \, g_Y^2 + g_Y^4}{24 \, \pi \, m_\chi^2} \,,
\label{eq:cross}
\end{equation}
where $g_2$ and $g_Y$ are the $SU(2)_L$ and $U(1)_Y$ gauge couplings, respectively. This interaction is strong enough to keep the Dark Matter particles in thermal equilibrium with the thermal bath, which are therefore produced through the standard freeze-out mechanism~\cite{Gondolo:1990dk}. In this framework, the Dark Matter relic abundance is given by~\cite{Giudice:2000ex}
\begin{equation}
\Omega_\chi h^2 \simeq 7.3\times10^{-11} \frac{1}{g^{1/2}_{*}\left(T_{\rm F,std}\right)}\frac{{\rm GeV^{-2}}}{\left< \sigma v \right> x^{-1}_{\rm F,std}} \,,
\label{eq:DMrelicSTD}
\end{equation}
where $g_{*}$ denotes the relativistic degrees of freedom of the thermal bath at the freeze-out temperature $T_{\rm F,std} = m_\chi / x_{\rm F,std}$ that is approximately given by the equation
\begin{equation}
x_{\rm F,std} \simeq \ln \left[0.038 \frac{g_\chi \, m_\chi \, M_{\rm Pl}\, x^{1/2}_{\rm F,std}}{g^{1/2}_{*}\left(T_{\rm F,std}\right)}\left< \sigma v \right>\right]\,,
\label{eq:xfstd}
\end{equation}
with $g_\chi = 6$ and $M_{\rm Pl} = 1.22\times10^{19}$~GeV is the Planck mass. By comparing Eq.~\eqref{eq:DMrelicSTD} with its experimental value provided by the Planck Collaboration $\Omega_{\rm DM} h^2 = 0.1186\pm0.0020$~\cite{Ade:2015xua}, we find that the model predicts a Dark Matter mass of about 2~TeV for which we have $\left<\sigma v\right> \simeq 2.8\times 10^{-26}~{\rm cm^3/s}$. Larger values for the Dark Matter mass would lead to larger values for the relic abundance implying an overclosure of the Universe. Indeed, from Eq.s~\eqref{eq:cross}~and~\eqref{eq:DMrelicSTD} it follows that the Dark Matter relic abundance is proportional to the square of the Dark Matter mass. Such expressions hold only if the Universe is dominated by radiation during the epoch of Dark Matter freeze-out. This is true for the standard scenario where it is generally assumed that the reheating temperature of the Universe, hereafter denoted as $T_{\rm RH}$, is much larger than the freeze-out temperature $T_{\rm F}$. However, the thermal history of the Universe before the Big Bang Nucleosynthesis is practically unknown and the reheating temperature could be as low as about 4~MeV~\cite{deSalas:2015glj}.

In scenarios with very low reheating~\cite{Lyth:1995ka,Cohen:2008nb,Fornengo:2002db,Gelmini:2006pq,Berlin:2016vnh,Berlin:2016gtr}, if $T_{\rm RH} \leq T_{\rm F}$, the equations~\eqref{eq:DMrelicSTD}~and~\eqref{eq:xfstd} become~\cite{Giudice:2000ex}
\begin{equation}
\Omega_\chi h^2 \simeq 2.3\times10^{-11} \frac{g^{1/2}_{*}\left(T_{\rm RH}\right)}{g_{*}\left(T_{\rm F,rh}\right)}\frac{T^3_{\rm RH}\,{\rm GeV^{-2}}}{m_\chi^3\,\left< \sigma v \right> x^{-4}_{\rm F,rh}} \,,
\label{eq:DMrelicRH}
\end{equation}
with
\begin{equation}
x_{\rm F,rh} \simeq \ln \left[0.015 \frac{g_\chi \, g^{1/2}_{*}\left(T_{\rm RH}\right)}{g_{*}\left(T_{\rm F,rh}\right)} \frac{M_{\rm Pl}\,T^2_{\rm RH}\, x^{5/2}_{\rm F,rh}}{m_\chi}\left< \sigma v \right>\right]\,.
\end{equation}
Therefore, for a given Dark Matter mass the correct relic abundance can be achieved by assuming a particular value for the reheating temperature. In particular, by numerically solving such equations in the range $10~{\rm TeV}\leq m_\chi \leq 10^4~{\rm TeV}$, we obtain the following approximated relation between the Dark Matter mass and the reheating temperature that provides the exact today's amount of Dark Matter in the Universe:
\begin{equation}
T_{\rm RH} \simeq 660 \left(\frac{m_\chi}{100~{\rm TeV}}\right)^{1/2}{\rm GeV} \,.
\end{equation}
For example, the correct Dark Matter relic abundance of Dark Matter particles with a mass of 100~TeV is achieved by assuming a reheating temperature of about 660~GeV.

A reheating of the Universe can be achieved by including at least an additional long-lived unstable particle $\phi$ that decays into radiation at a time of the order of its lifetime $\Gamma_\phi$. These non-relativistic particles could dominate the energy density of Universe providing a matter-dominated expansion instead of a radiation-dominated one. Then, the decays of such particles into relativistic particles of the thermal bath reheat the Universe at a temperature $T_{\rm RH}$ defined through the equation
\begin{equation}
\Gamma_\phi = \sqrt{\frac{4\pi^3\,g_{*}\left(T_{\rm RH}\right)}{45}}\frac{T^2_{\rm RH}}{M_{\rm Pl}} \,.
\end{equation}
This result generally occurs during the inflation where the particle $\phi$ is identified with the inflaton. Hence, it is reasonable to assume that different reheat events could occur after inflation, especially in the presence of additional highly decoupled sectors~\cite{Berlin:2016vnh,Berlin:2016gtr}.

\section{Compatibility with IceCube and Fermi-LAT}

Having defined the model that provides a viable Dark Matter candidate that decays into a neutrino line only, we can now proceed to analyze its compatibility with the IceCube 6-year HESE data~\cite{Aartsen:2017mau}. In particular, in our scenario, the TeV-PeV extraterrestrial neutrino flux has two contributions, a power-law accounting for neutrinos produced by standard astrophysical flux and a neutrino flux originated by Dark Matter decays. Hence, the total differential extraterrestrial neutrino flux for each neutrino flavor $\alpha$ is given by
\begin{equation}
\frac{{\rm d}\phi_\alpha}{{\rm d}E_\nu {\rm d}\Omega} = \frac{{\rm d}\phi^{\rm Astro}_\alpha}{{\rm d}E_\nu {\rm d}\Omega} + \frac{{\rm d}\phi^{\rm DM}_\alpha}{{\rm d}E_\nu {\rm d}\Omega}\,.
\label{eq:tot_flux}
\end{equation}
The astrophysical contribution is parametrized by a power-law
\begin{equation}
\frac{{\rm d}\phi^{\rm Astro}_\alpha}{{\rm d}E_\nu {\rm d}\Omega} = \phi^{\rm Astro}_0 \left(\frac{E_\nu}{100~{\rm TeV}}\right)^{-\gamma} \,,
\end{equation}
where $\phi^{\rm Astro}_0$ is the normalization of the flux at 100~TeV and $\gamma$ is the spectral index. According to the 8-year through-going muon neutrinos data and the recent measurements related to the blazar TXS~0506+056, we fix the spectral index to the benchmark value $\gamma=2.2$. Therefore, only the normalization $\phi^{\rm Astro}_0$ is taken as free parameter. We note that the astrophysical flux is isotropic and independent on the neutrino flavor $\alpha$.

The Dark Matter neutrino flux in Eq.~\eqref{eq:tot_flux} is instead given by the sum of a Galactic (G) contribution and an Extragalactic (EG) one. Hence, taking into account the neutrino oscillations through mixing probabilities $P_{\alpha\beta}$~\cite{Chianese:2016kpu}, we have
\begin{equation}
\frac{{\rm d}\phi^{\rm DM}_\alpha}{{\rm d}E_\nu {\rm d}\Omega} = \sum_{\beta} P_{\alpha\beta} \left[ \frac{{\rm d}\phi^{\rm G}_\beta}{{\rm d}E_\nu {\rm d}\Omega} + \frac{{\rm d}\phi^{\rm EG}_\beta}{{\rm d}E_\nu {\rm d}\Omega} \right]\,,
\label{eq:flux}
\end{equation}
with
\begin{eqnarray}
\frac{{\rm d}\phi^{\rm G}_\beta}{{\rm d}E_\nu {\rm d}\Omega} & = & \frac{1}{4\pi \, m_\chi \, \tau_\chi} \frac{{\rm d}N_\beta}{{\rm d}E_\nu} \int_0^\infty ds \, \rho_{\rm NFW}\left(s,\ell,b\right)\,,\\
\frac{{\rm d}\phi^{\rm EG}_\beta}{{\rm d}E_\nu {\rm d}\Omega}& = &  \frac{\Omega_\chi\rho_c}{4\pi \, m_\chi \, \tau_\chi} \int_0^\infty dz \,  \frac{1}{H\left(z\right)}\left.\frac{{\rm d}N_\beta}{{\rm d}E_\nu}\right|_{E\left(1+z\right)}\,.
\end{eqnarray}
Both contributions are inversely proportional to the Dark Matter mass $m_\chi$ and its total lifetime $\tau_\chi$. The quantity ${\rm d}N_\beta/{\rm d}E_\nu$ is the effective energy spectrum of neutrinos produced by each Dark Matter decay. It is given by the sum of the energy spectra of each Dark Matter decay channel (different neutrino flavors) multiplied by the corresponding branching ratio. To evaluate such a quantity, we use the tables provided by Ref.~\cite{Cirelli:2010xx}.\footnote{The energy spectra have been extrapolated up to a Dark Matter mass of 10 PeV by using the same approach discussed in Ref.~\cite{Chianese:2016kpu}.} The Galactic flux is proportional to the integral over the line-of-sight $s$ of the Dark Matter halo density profile of the Milky Way $\rho_{\rm NFW}$, which we assume to be the standard Navarro-Frenk-White (NFW) distribution~\cite{Navarro:1995iw}. The Extragalactic flux is instead obtained by integrating over the redshift $z$ the energy spectrum evaluated at the energy $E\left(1+z\right)$ and divided by the Hubble parameter $H\left(z\right)$. For the cosmological parameters, we take the $\Lambda$CDM parameters provided by the Planck Collaboration~\cite{Ade:2015xua}. Finally, it is worth noticing that the Galactic term depends on the Galactic angular coordinates $\left(b,\ell\right)$ through the Dark Matter halo density profile, while the Extragalactic one is isotropic.

Hence, the two-component neutrino flux defined in Eq.~\eqref{eq:tot_flux} depends on three free parameters: the astrophysical flux normalization $\phi^{\rm Astro}_0$, the Dark Matter mass $m_\chi$ and its total lifetime $\tau_\chi$. For each choice of these three quantities, the expected number of neutrino events in a given energy bin $\left[E_i,E_{i+1}\right]$ of the IceCube HESE data sample is obtained as
\begin{equation}
N_i = \Delta t \int_{E_i}^{E_{i+1}} dE_\nu \int d\Omega \,\sum_\alpha \frac{{\rm d}\phi_\alpha}{{\rm d}E_\nu {\rm d}\Omega} A_\alpha\left(E_\nu,\Omega\right)\,,
\end{equation}
where $\Delta t = 2078$~days is the exposure time of the 6-years HESE data and the quantity $A_\alpha \left(E_\nu,\Omega\right)$ is the effective area of the IceCube detector per neutrino flavor $\alpha$~\cite{Aartsen:2013jdh}. In order to provide the allowed regions for the Dark Matter parameters, $m_\chi$ and $\tau_\chi$, the expected number of neutrinos has to be compared with the observed one in each energy bin. This is done by means of a binned multi-Poisson likelihood~\cite{Baker:1983tu}, whose expression is
\begin{equation}
\ln \mathcal{L} = \sum_i\left[n_i - N_i + n_i \ln\left(\frac{N_i}{n_i}\right)\right]\,,
\label{eq:likelihood}
\end{equation}
where $n_i$ is the observed number of neutrinos once the background events have been subtracted in each bin $i$. In particular, we consider only the conventional atmospheric background (atmospheric neutrinos and penetrating muons)~\cite{Honda:2006qj}, while the prompt atmospheric background (neutrinos produced by the charmed mesons decays)~\cite{Enberg:2008te} is assumed to be negligible, according to IceCube results reported in Ref.s~\cite{Aartsen:2017mau,Aartsen:2013eka,Aartsen:2014muf}.

\begin{figure}[t!]
\begin{center}
\includegraphics[width=0.65\textwidth]{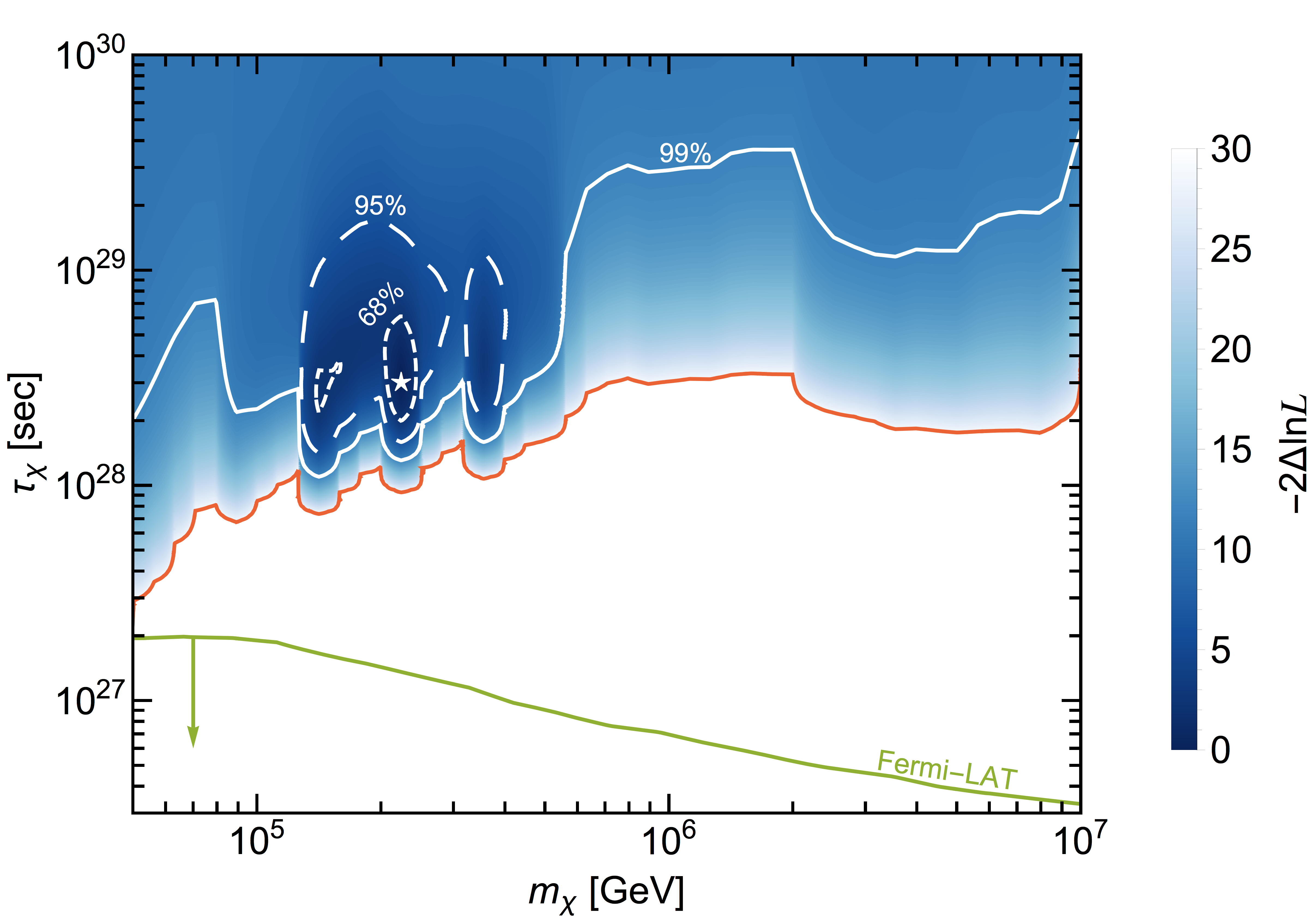}
\end{center}
\caption{\label{fig:ICdec}Profile of the likelihood in the plane $m_\chi$--$\tau_\chi$. The white contours delimit the regions at 68\% (short dashed), 95\% (long dashed) and 99\% (solid) C.L., while the red solid line bounds from above the regions excluded at more than $5\sigma$. The white star corresponds to the best-fit point. The green line represents the limit on the decaying Dark Matter neutrino line deduced by the Fermi-LAT data~\cite{Cohen:2016uyg}.}
\end{figure}

In Fig.~\ref{fig:ICdec}, we report the profile of the likelihood given in Eq.~\eqref{eq:likelihood} in the plane $m_\chi$--$\tau_\chi$.

The confidence contour levels (C.L.) have been obtained by considering the astrophysical flux normalization as nuisance parameter. According to the Wilks' theorem, the quantity $-2\Delta \ln\mathcal{L}$ follows a chi-squared distribution with two degrees of freedom. In the plot, the regions at 68\% (short dashed), 95\% (long dashed) and 99\% (solid) C.L. are delimited by the white contours. The region of Dark Matter parameters excluded at more than $5\sigma$ is bounded from above by the red solid line. Moreover, the green solid line shows instead the constraint on our model deduced by the Fermi-LAT gamma-rays measurements~\cite{Cohen:2016uyg}. Indeed, even though our Dark Matter candidate decays into a neutrino line only, electrons/positrons and gamma-rays are produced as well through the radiative corrections that induce $W$/$Z$ bremsstrahlung. Moreover, it is worth observing that, in addition to the neutrinophilic decays, our model also allows for Dark Matter annihilations into Standard Model particles, especially into electroweak gauge bosons. However, the present gamma-rays constraints on these annihilation channels are not relevant since, according to Eq.~\eqref{eq:cross}, the thermally average cross-section is very small for a Dark Matter mass larger than 1~TeV~\cite{Fermi-LAT:2016uux,Abeysekara:2017jxs}. Finally, we note that our results are compatible with the corresponding limits provided by the IceCube Collaboration~\cite{Aartsen:2018mxl}.

\begin{figure}[t!]
\begin{center}
\includegraphics[width=0.6\textwidth]{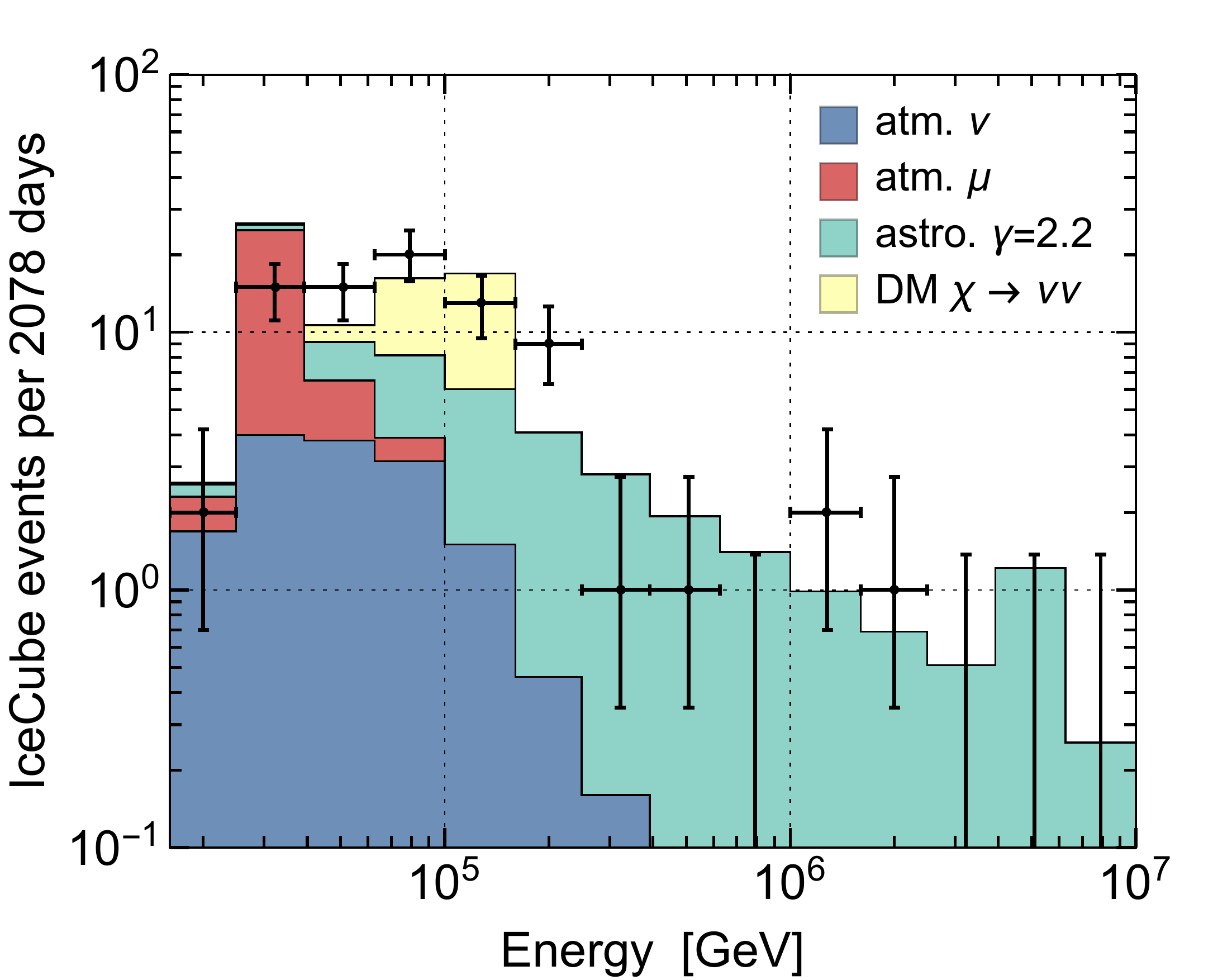}
\end{center}
\caption{\label{fig:ICbest}Numbers of neutrino events as a function of the neutrino energy after 2078 days of data-taking, for the best-fit two-component flux. The astrophysical contribution (green color) is a power-law with a spectral index 2.2 and normalization $\phi^{\rm Astro}_0 = 0.4\times 10^{-18}~{\rm GeV^{-1}cm^{-2}sec^{-1}sr^{-1}}$. The Dark Matter contribution (yellow color) corresponds to a Dark Matter mass $m_\chi\simeq220~{\rm TeV}$ and a lifetime $\tau_\chi \simeq 3 \times 10^{28}~{\rm sec}$.}
\end{figure}

The maximum of the likelihood (best-fit point shown with a white star) has been found for the following values: $\phi^{\rm Astro}_0 = 0.4\times 10^{-18}~{\rm GeV^{-1}cm^{-2}sec^{-1}sr^{-1}}$, $m_\chi\simeq220~{\rm TeV}$ and $\tau_\chi \simeq 3 \times 10^{28}~{\rm sec}$. In Fig.~\ref{fig:ICbest} we show the number of neutrino events as a function of the neutrino energy for the best-fit two-component neutrino flux. The different contributions are shown with different colors: the atmospheric background in blue (neutrinos) and in red (penetrating muons), the astrophysical power-law in green, and the decaying Dark Matter component in yellow. It is worth observing that, even though the neutrino flux provided by a neutrino line is mainly peaked at $E_\nu = m_\chi /2$, the best-fit Dark Matter component provides a significant contribution to two energy bins around 100~TeV. Indeed, the maximum of the likelihood occurs when the half of the Dark Matter mass is exactly between the two energy bins. This also explains the peculiar behavior of the likelihood profile: the dips of the contours displayed in Fig.~\ref{fig:ICdec} are indeed in correspondence of the transition between two energy bins of the IceCube HESE data sample. A more detailed likelihood analysis would require the unbinned data set that is not public so far.

\begin{figure}[t!]
\begin{center}
\includegraphics[width=0.6\textwidth]{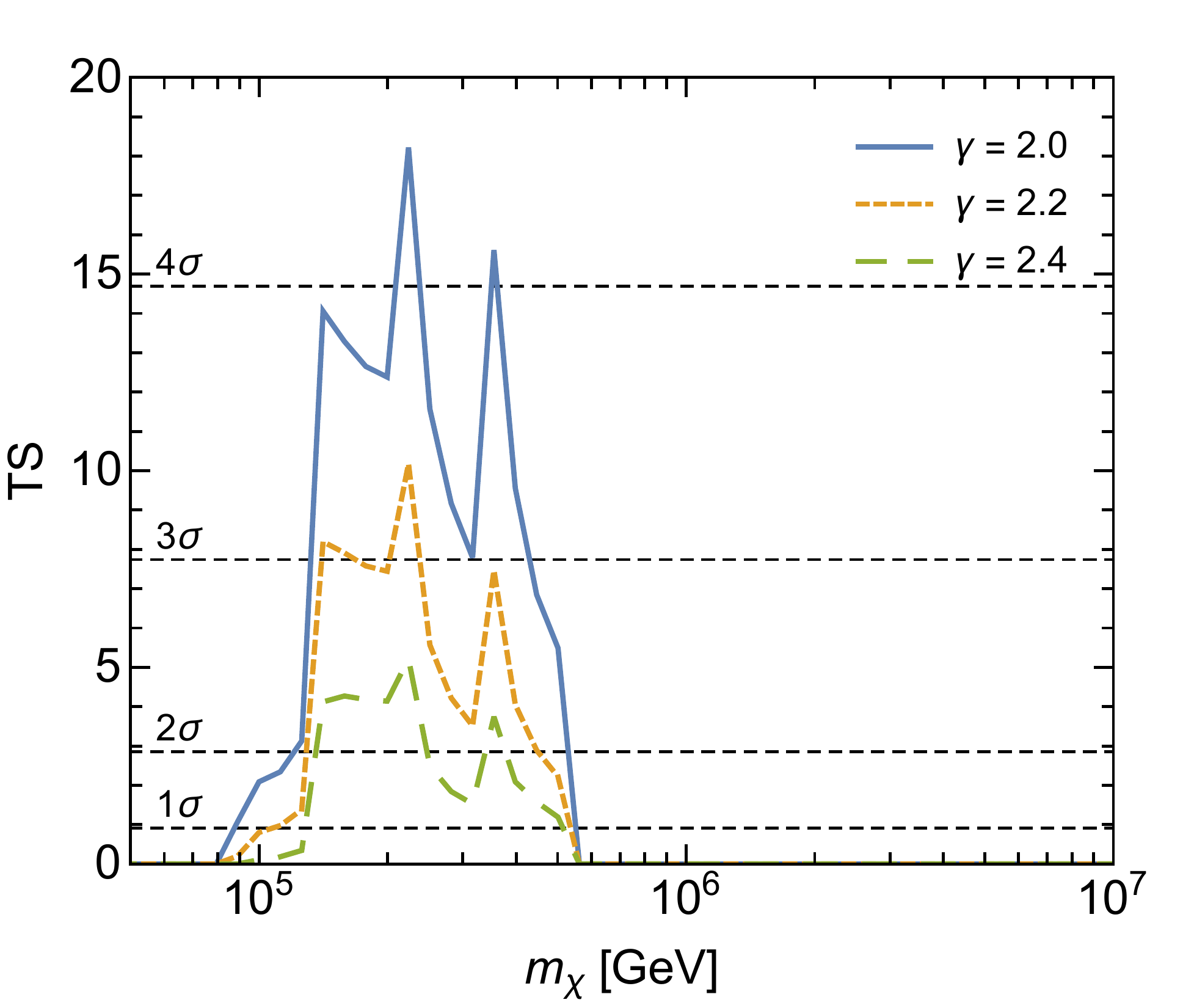}
\end{center}
\caption{\label{fig:TS}Significance of the Dark Matter neutrino line signal as a function of the Dark Matter mass, for different values of the spectral index.}
\end{figure}

Moreover, we statistically quantify how the fit is improved by including a Dark Matter neutrino line component on top of a hard astrophysical power-law. This is done by means of the following test statistics
\begin{equation}
{\rm TS}\left(m_\chi\right) = -2 \ln \frac{\mathcal{L}\left(\phi_0^{\rm Astro},\tau_\chi\to\infty\right)}{\mathcal{L}\left(\phi_0^{\rm Astro},\tau_\chi,m_\chi\right)} \,.
\label{eq:TS}
\end{equation}
For each value of the Dark Matter mass, we compare the maximum likelihood of the null-hypothesis (astrophysical power-law flux only corresponding to $\tau_\chi\to\infty$) with respect to the one of the signal hypothesis (two-component neutrino flux). According to the Wilks' theorem, this test statistics follows the distribution $\frac12 \delta\left({\rm TS}\right) + \frac12 \chi^2_1\left({\rm TS}\right)$ through which we compute the significance of the Dark Matter signal (see Fig.~\ref{fig:TS}). Hence, we find that a Dark Matter signal is favored at more than $3\sigma$ with respect to the only astrophysical power-law with spectral index $\gamma = 2.2$.

Finally, we comment on what happens if one relaxes the assumption of fixing the spectral index of the astrophysical power-law to the benchmark value 2.2. In particular, Fig.~\ref{fig:TS} shows that the significance of the Dark Matter signal ranges from 2$\sigma$ to 4$\sigma$ as a function of the spectral index. Here, the range $2.0\div2.4$ just covers the best-fit values of different analyses of the TXS 0506+056 blazar and it is also in agreement with the through-going muon neutrinos data sample. On the other hand, if one takes the spectral index as a completely free parameter, the best-fit two-component neutrino flux would be a soft power-law (spectral index larger than 3.0) plus a Dark Matter contribution that explains the three PeV neutrino events. Hence, the best-fit value for the Dark Matter mass would be $m_\chi = \mathcal{O}(\rm PeV)$. We note that the features of a Dark Matter neutrino line flux are indeed in perfect agreement with observation of the three PeV neutrinos, that are the most statistically events having with no background. However, we stress once again that a soft power-law is disfavor by astrophysical observations.

\section{Conclusions}

The recent measurement of the neutrino flux produced by the blazar TXS~0506+056 and the IceCube analysis on the 8-year through-going muon neutrinos from the Northern Sky suggest the presence of a hard power-law component in the extraterrestrial TeV-PeV neutrino flux. This is slightly in contrast with IceCube 6-year HESE data that prefer instead a soft power-law neutrino flux. Such a tension could indicate that the observed extraterrestrial TeV-PeV neutrino flux is given by the sum of different contributions. One component could be intriguingly originated by heavy decaying Dark Matter particles. However, most of Dark Matter models predict very large gamma-rays flux as well and, therefore, are very constrained by Fermi-LAT observations. Such gamma-rays constraints can be circumvented in case of a neutrinophilic Dark Matter.

In the present paper, we have investigated a model of neutrinophilic Dark Matter particles that decay into a pure neutrino line only. The role of Dark Matter is played by the neutral component of a new $SU(2)_L$-triplet scalar with hypercharge $Y=+1$. The neutrinophilic nature of Dark Matter particles is achieved by introducing a new symmetry, like a global $U(1)$, under which only leptons and the new field are charged. Such a symmetry forbids the other potential Dark Matter decay channels that are induced by the couplings with the Higgs field. Since the symmetry must remain preserved, the only way to give mass to active neutrinos is through the Dirac Yukawa coupling once right-handed neutrinos are introduced in the model. Hence, as a consequence, the active neutrinos are Dirac particles.

In the addition to the decays into neutrinos, the Dark Matter has also electroweak interactions that are responsible for its production in the early Universe through the standard freeze-out mechanism. According to the WIMP paradigm, the Dark Matter mass is fixed to be about 2~TeV in order to achieve the correct Dark Matter relic abundance. However, in order to allow for heavier Dark Matter particles, we have considered a very low-reheating temperature of the Universe, smaller than the Dark Matter mass. In particular, we have obtained an approximated numerical relation between the reheating temperature and the Dark Matter mass providing the correct today's amount of Dark Matter.

Finally, we have analyzed the regions of the parameters space that are allowed by IceCube and Fermi-LAT. We have performed a likelihood statistical analysis to fit the IceCube 6-year HESE data with a two-component neutrino flux:  an astrophysical power-law with spectral index 2.2 and the contribution of the Dark Matter neutrino line. We have found as best-fit a Dark Matter mass of about 200 TeV and a Dark Matter lifetime of about $3\times 10^{28}$~sec. Such a scenario predicts a reheating temperature of the Universe of about 1~TeV, well above the limit provided by Big Bang Nucleosynthesis.

\section*{Acknowledgements}

The work of GM and SM is partially supported by Istituto Nazionale di Fisica Nucleare (INFN) through the �Theoretical Astroparticle Physics� (TAsP) project. MC acknowledges financial support from the STFC Consolidated Grant P000711/1. EP acknowledges financial support from  DGAPA-PAPIIT IN107118, the German-Mexican research collaboration grant SP 778/4-1 (DFG) and 278017 (CONACyT) and PIIF UNAM.

\end{document}